\documentclass[usenatbib]{mn2e}
\usepackage{graphicx}

\title[The T$_{\rm e}$ determination]
      {On the electron temperature determination in high-metallicity 
      H\,{\sc ii} regions}

\author[L.S. Pilyugin]
       {Leonid S. Pilyugin\\
        Main Astronomical Observatory
        of National Academy of Sciences of Ukraine,
        27 Zabolotnogo str., 03680 Kiev, Ukraine \\
        {\it pilyugin@mao.kiev.ua}\\} 

\date{Accepted 2006 November 21. Received 2006 November 21; 
                              in original form 2006 October 19}

\begin{document}

\maketitle

\begin{abstract}
The problem of determination of the electron temperature t$_2$ in the O$^{+}$ 
zone of high-metallicity H\,{\sc ii} region was examined. It was shown that 
the ratio of nebular to auroral nitrogen line intensities, which is an 
indicator of the electron temperature t$_2$, can be expressed in terms of 
the nebular line intensities of oxygen. This solves 
the problem of the determination of the electron temperature t$_2$, since 
the oxygen nebular lines are strong and, consequently, are readily observable.  
A relation between electron temperatures in the O$^{+}$ and O$^{++}$ zones 
in high-metallicity H\,{\sc ii} regions was studied. 
It was found that there is no one-to-one correspondance between t$_2$ 
and t$_3$ temperatures. Instead the t$_2$ -- t$_3$ relation is dependent on 
excitation parameter. 
\end{abstract}


\section{Introduction}

The oxygen abundance is one of the fundamental characteristics of a galaxy.
Accurate oxygen abundances are mandatory in investigations of different 
aspects of the formation and evolution of galaxies.
Accurate oxygen abundances in H\,{\sc ii} regions can be derived via the 
classic T$_{\rm e}$ method. In the first step, the electron temperature $t_3$ 
within the [O\,{\sc iii}] zone and the electron temperature $t_2$ within the 
[O\,{\sc ii}] zone are determined. Then the abundance is derived using the 
equations linking the ionic abundances to the measured line intensities and 
electron temperature. 
(The usual notation (O/H)$_{\rm T_e}$ for the oxygen abundance 
derived with the T$_{\rm e}$ method will be used throughout the paper.)
The ratio of nebular to auroral oxygen line intensities 
Q$_{\rm OIII}$ = [O\,{\sc iii}]$\lambda 4959+\lambda 5007$/[O\,{\sc iii}]$\lambda 4363$ 
is used for the t$_3$ determination. The ratio of 
nebular to auroral oxygen line intensities Q$_{\rm OII}$ = 
[O\,{\sc ii}]$\lambda$3727/[O\,{\sc ii}]$\lambda$7320+$\lambda$7330
or the ratio of nebular to auroral nitrogen line intensities Q$_{\rm NII}$ = 
[N\,{\sc ii}]$\lambda 6548+\lambda 6584$/[N\,{\sc ii}]$\lambda 5755$ 
are used for the t$_2$ determination. Unfortunately, the auroral lines are 
faint and drop below detectability in the spectra
of high-metallicity H\,{\sc ii} regions. This prevents the application of the 
T$_{\rm e}$ method to high-metallicity H\,{\sc ii} regions. 

Is there another way to estimate electron temperature in H\,{\sc ii} region 
where the faint auroral lines are not detected?
The law of energy conservation for free electrons \citep[e.g.][]{sobolev} shows 
that the electron temperature in a nebula is mainly defined by the hardness of 
the ionizing radiation and by the emission in the forbidden lines. Since the 
excitation parameter P is an indicator of the hardness of the ionizing 
radiation, one can expect that the excitation parameter P coupled 
with the measured intensities of the nebular oxygen lines can be used to 
estimate the electron temperature 
in H\,{\sc ii} regions. We have suggested that the parametric calibration 
links the oxygen abundance to the measured nebular oxygen line intensities 
and the excitation parameter P (P calibration or P method) \citep{lcal,hcal,vybor}.
(The oxygen abundance derived with the P calibration will be referred to 
as (O/H)$_{\rm P}$.) 
Comparisons between (O/H)$_{\rm T_e}$ and (O/H)$_{\rm P}$ abundances  
in H\,{\sc ii} regions with direct measurement of the electron temperature 
as well as the comparison between radial distributions of (O/H)$_{\rm T_e}$ and 
(O/H)$_{\rm P}$ abundances in the disks of a few well studied spiral galaxies 
show that (O/H)$_{\rm P}$ abundances are in agreement with (O/H)$_{\rm T_e}$ 
abundances \citep{m101,mwg,pilvilcon04,pilyuginthuan05}.
This indirectly confirms that the physical conditions in the H\,{\sc ii} 
region can be estimated via the nebular oxygen line intensities and the 
excitation parameter P.
(Or one may also say that the physical conditions in the H\,{\sc ii} 
region can be estimated via the nebular oxygen line intensities only, 
since the excitation parameter P is expressed in terms of these lines.)  
If this is the case, then one can expect 
that the diagnostic line ratios can be expressed in terms of the intensities 
of the oxygen nebular lines. Indeed, it has been found empirically that 
there is a relationship (the ff relation) between auroral 
[OIII]$\lambda 4363$ and nebular oxygen line intensities in spectra of H\,{\sc ii} 
regions \citep{ff,pilyuginetal06}. In other words, the diagnostic line ratio 
Q$_{\rm OIII}$, used for the determination of the electron temperature 
within the [O\,{\sc iii}] zone, can be expressed in terms of the nebular 
oxygen line intensities. 
It has also been suggested that the electron temperature $t_2$ within the 
[O\,{\sc ii}] zone is related to the electron temperature $t_3$ 
within the [O\,{\sc iii}] zone 
\citep{campbelletal86,pageletal92,izotovetal97,deharvengetal00,oeyshields00,pvt06}. 
Then one can expect that the diagnostic line ratio 
used for the determination of the electron temperature within the [O\,{\sc ii}] 
zone, can also be expressed in terms of the nebular oxygen line intensities. 
The goal of this paper is to establish this relation. 
 
The paper is organized as follows. 
A relation between the ratio of nebular to auroral nitrogen line intensities 
and oxygen nebular line intensities, Q$_{\rm NII}$ = f(R$_2$,P), 
is derived in Section 2. 
A relation between electron temperatures t$_2$ and t$_3$ is discussed 
in Section 3. 
We summarize our conclusions in Section 4.

We will be using the following notations throughout the paper:
R$_2$ = I$_{{\rm [OII] \lambda 3727+ \lambda 3729}}$/I$_{{\rm H_{\beta} }}$,
R$_3$ = I$_{{\rm [OIII] \lambda 4959+ \lambda 5007}}$/I$_{{\rm H_{\beta} }}$,
R = I$_{{\rm [OIII] \lambda 4363}}$/I$_{{\rm H_{\beta} }}$. 
With these definitions, the excitation parameter P can be 
expressed as: P = R$_3$/(R$_3$+R$_2$). 

%------------------------------------------------------------
\section{The relation Q$_{\rm NII}$ = \lowercase{f}(R$_2$,P)}
%-------------------------------------------------------------

%====================================Fig  1    f - f
\begin{figure}
\resizebox{1.00\hsize}{!}{\includegraphics[angle=000]{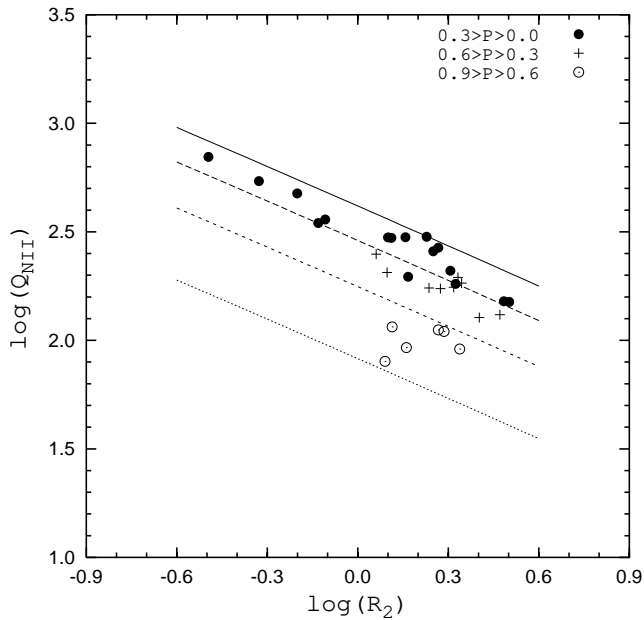}}
\caption{Plot of log(Q$_{\rm NII}$) as a function of log(R$_{2}$) for a sample of 
H\,{\sc ii} regions. 
The filled circles are H\,{\sc ii} regions with 0 $<$ P $<$ 0.3. 
The plus signs are those with 0.3 $<$ P $<$ 0.6. The open circles 
are those with P $>$ 0.6. 
The relations corresponding to Eq.(\ref{equation:ff}) for different values 
of the excitation parameter are shown by the solid (P = 0.0),  
long-dashed  (P = 0.3), short-dashed  (P = 0.6), 
and dotted (P = 0.9) lines. 
}
\label{figure:ff}
\end{figure}

%====================================Fig  2    f - df
\begin{figure}
\resizebox{1.00\hsize}{!}{\includegraphics[angle=000]{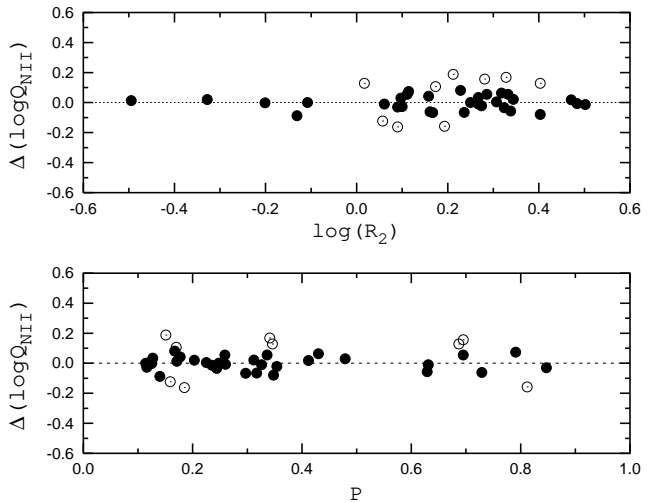}}
\caption{Plot of $\Delta$(logQ$_{\rm NII}$) as a function of nebular oxygen line 
intensity log(R$_{2}$) ({\it top panel}) and of excitation parameter P 
({\it bottom panel}). 
The filled circles are H\,{\sc ii} regions used in deriving the final relation. 
The open circles are the other H\,{\sc ii} regions. 
}
\label{figure:f-df}
\end{figure}

The electron temperature $t_2$ within the O$^+$ zone in H\,{\sc ii} regions
can be derived from the diagnostic line ratio Q$_{\rm OII}$ or from the 
diagnostic line ratio Q$_{\rm NII}$.  
It is believed that the measured Q$_{\rm NII}$ results in more reliable t$_2$ value. 
Starting from our basic idea that the diagnostic line ratio can be expressed 
in terms of the strong oxygen nebular line intensities, we will search for 
a relationship of the type Q$_{\rm NII}$ = f(R$_2$,P) here.  
It should be noted that the ratio of nebular to auroral nitrogen line 
intensities Q$_{\rm NII}$ = 
[N\,{\sc ii}]$\lambda 6548+\lambda 6584$/[N\,{\sc ii}]$\lambda 5755$ 
is an indicator of the electron temperature t$_2$ and therefore this ratio 
depends on the electron temperature but not on the nitrogen abundance.  

Recent measurements of the oxygen and nitrogen line intensities in 
high-metallicity (12+logO/H $>$ 8.25) H\,{\sc ii} regions were taken from
\citet{castellanosetal02,luridianaetal02,peimberta03,kennicuttetal03,
tsamisetal03,bresolinetal04} and \citet{bresolinetal05}. 
These spectroscopic data (40 data points) form the basis of 
the present study.

The value of Q$_{\rm NII}$ is shown as a function of the intensity of the  
nebular oxygen line R$_{2}$ in Fig.~\ref{figure:ff} (the points used in the 
determination of the final relation are shown). 
The filled circles are H\,{\sc ii} regions with 0 $<$ P $<$ 0.3. 
The plus signs are those with 0.3 $<$ P $<$ 0.6. The open circles 
are those with P $>$ 0.6. Inspection of Fig.~\ref{figure:ff} 
shows that the value Q$_{\rm NII}$ is linked to the intensity of the nebular 
oxygen line R$_{2}$ through a relation of the type
\begin{eqnarray}
\log ({\rm Q_{NII}}) & = &   a_0 + a_1\,\log ({\rm R_2}) + 
a_2 \, [\log ({\rm R_2})]^2 
\nonumber  \\
       & + & b_1 \, \log (1-{\rm P})  +  b_2 \, [\log (1-{\rm P})]^2 .
\label{equation:ffa}
\end{eqnarray}
Using a sample of H\,{\sc ii} regions with measurements of oxygen and nitrogen 
line intensities, the values of the coefficients in Eq.(\ref{equation:ffa}) 
can be derived. The values of the coefficients in Eq.(\ref{equation:ffa}) 
are derived by using an iteration procedure. In the first step, the relation is 
determined from all data using the least-square method. Then, the point with  
the largest deviation is rejected, and a new relation is derived. The iteration 
procedure is pursued until two successive relations have all their coefficients 
differing by less 0.001 and the absolute value of the largest deviation is 
less than 0.1 dex. The following relation was obtained
\begin{eqnarray}
\log ({\rm Q_{NII}}) & = &   2.619 - 0.609\,\log ({\rm R}_{2}) - 
0.010 \, [\log ({\rm R}_{2})]^2 
\nonumber  \\
       & + &
1.085 \, \log (1-{\rm P})  +  0.382 \, [\log (1-{\rm P})]^2 .
\label{equation:ff}
\end{eqnarray}
The final relation was obtained using 31 data points out of an original set 
of 40 points. The relations corresponding to Eq.(\ref{equation:ff}) for 
different values of the excitation parameter are shown in Fig.~\ref{figure:ff} 
by the solid (P = 0.0), long-dashed  (P = 0.3), short-dashed  (P = 0.6), 
and dotted (P = 0.9) lines. 

The differences $\Delta$(logQ$_{\rm NII}$) between measured values of 
logQ$_{\rm NII}^{obs}$ and those logQ$_{\rm NII}^{cal}$ computed with 
obtained relation are shown as a function of oxygen line intensity in 
Fig.\ref{figure:f-df} (top panel) and as a function of excitation 
parameter P (bottom panel). The points used in the determination of the final 
relation are shown by filled circles. The line is the linear best fit to those 
data derived through the least squares method. For comparison, we also show by 
open circles the objects that have been rejected by the iteration procedure. 
Fig.~\ref{figure:f-df} shows that the deviations  do not show a correlation 
either with oxygen line intensity or with excitation parameter. 

Examination of Fig.\ref{figure:ff} and Fig.\ref{figure:f-df} shows that 
Eq.(\ref{equation:ff}) gives a satisfactory fit to the  observational data. 
It should be noted 
however that the particular form of the analytical expression adopted here 
may be questioned. We have chosen a simple form, Eq.(\ref{equation:ffa}),  
but perhaps a more complex expression may give a better fit to the data. 
Furthermore, the error in measurements of the line [NII]$\lambda 5755$ is in 
excess of 20\% (as indicated in original papers) in a number of calibrating 
H\,{\sc ii} regions. Perhaps more accurate measurements  may result in a more 
precise relation. Clearly, high-precision measurements of oxygen and nitrogen 
lines in spectra of H\,{\sc ii} regions are needed to check the derived 
relation\footnote{After this study was carried out, new high-precision 
measurements of oxygen and nitrogen lines in the spectrum of H\,{\sc ii} region 
H1013 of the spiral galaxy M101 were 
published by \citet{bresolin06}. This provides an additional possibility to
check the derived relations. The measured intensity of [NII]5755 is 
0.0059$\pm$0.0003 on a scale where I$_{H_{\beta}}$=1, and Eq.(\ref{equation:ff}) results in 0.00600. 
The measured intensity of [OIII]4363 is 
0.0024$\pm$0.0003, and Eq.(\ref{equation:ffo}) results in 0.00267. Thus, 
the differences between measured line intensities and those predicted by our 
relation is not in excess of the reported uncertainty of the measurement 
both for [NII]5755 and [OIII]4363 lines.}.  

Thus, Eq.(\ref{equation:ff}) confirms our basic idea that the diagnostic line 
ratio, which is an indicator of the electron temperature t$_2$, can be 
expressed in terms of the oxygen nebular line intensities. Since the oxygen 
nebular lines are strong and, consequently, are easily observable, 
Eq.(\ref{equation:ff}) solves the problem of the electron temperature 
determination in O$^+$ zone of H\,{\sc ii} region where the faint auroral 
nitrogen line is not detected. It should be emphasized that the obtained 
relation Q$_{\rm NII}$ = f(R$_2$,P) is a purely empirical relation in the sense 
that this is the relation between directly measured values and, consequently, 
this relation is not based on any assumption. 

%============================================================
\section{The \lowercase{t$_2$} -- \lowercase{t$_3$} relation}
%============================================================

%---------------------------------------------------
\subsection{One-dimensional t$_2$ -- t$_3$ relation}
%---------------------------------------------------

Here we will consider the t$_2$ -- t$_3$ relation for our sample of H\,{\sc ii} 
regions. The measured Q$_{\rm NII}$ values are used to determine the electron 
temperatures t$_2$. To establish an expression that links the electron temperature
t$_2$ to the value of the Q$_{\rm NII}$, we have performed a five-level-atom 
calculation using recent atomic data. The Einstein coefficients for spontaneous 
transitions and the energy levels for five low-lying levels were taken from 
\citet{galavisetal97}. The effective cross sections or effective collision 
strengths for electron impact were taken from \citet{hudsonbell05}. 
The effective cross sections are continuous functions of temperatures and 
are tabulated by \citet{hudsonbell05} at fixed temperatures. The actual 
effective cross sections for a given electron temperature used here were derived from 
two-order polynomial fits of the data from \citet{hudsonbell05} as a function 
of temperature. The expression for the H$_{\beta}$ emissivity was taken from 
\citet{aller84}. The five-level-atom solution for ion N$^+$ results in the 
following simple expression for the determination of t$_2$
\begin{equation}
t_2 = \frac{1.111}{\log (Q_{\rm NII})  - 0.892 - 0.144 \, \log (t_2) + 0.023 \, t_2}.
\label{equation:t2n}
\end{equation}
The majority of extragalactic H\,{\sc ii} regions are in the low-density regime 
\citep{zkh,bresolinetal05}. 
Therefore the electron density n$_{\rm e}$ = 100 cm$^{-3}$ was adopted for 
the five-level-atom calculation, and for this reason there is no n$_{\rm e}$-term 
in Eq.(\ref{equation:t2n}). 

The electron temperature t$_3$ is derived from the Q$_{\rm OIII}$ = R$_{3}$/R 
ratio. The relation between t$_3$ and Q$_{\rm OIII}$ is based on the 
five-level-atom solution. 
The Einstein coefficients for spontaneous transitions and the energy levels 
were taken from \citet{galavisetal97}, while the effective cross sections for electron 
impact were taken from \citet{aggarwal99}. Again, the actual 
effective cross sections for a given electron temperature were derived from 
two-order polynomial fits of the data from \citet{aggarwal99} as a function 
of temperature. The five-level-atom solution for ion O$^{++}$ results in
following simple expression for the determination of t$_3$
\begin{equation}
t_3 = \frac{1.432}{\log (Q_{\rm OIII})  - 0.875 -0.025 \, \log (t_3) - 0.020 \, t_3}.
\label{equation:t3o}
\end{equation}
It should be noted that the electron temperatures t$_3$ derived from 
Eq.(\ref{equation:t3o}) are close to those derived from analogous expressions given 
in \citet{pageletal92} and in \citet{izotovetal06}.

%====================================Fig  3    t2 - t3 single
\begin{figure}
\resizebox{1.00\hsize}{!}{\includegraphics[angle=000]{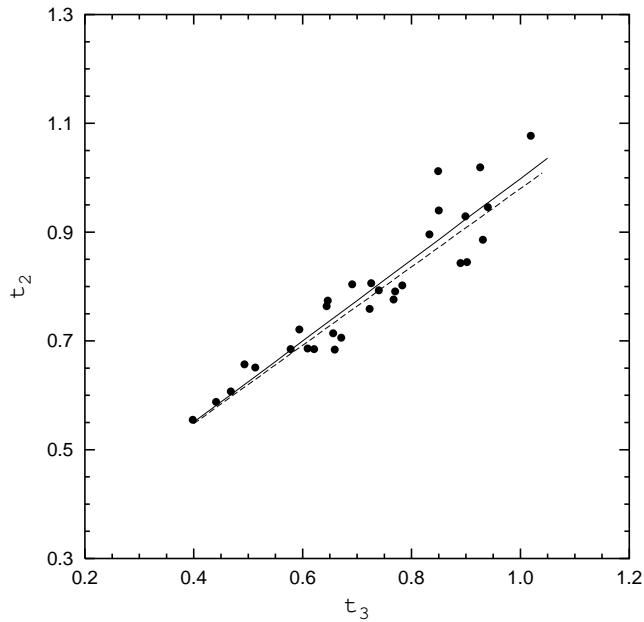}}
\caption{The t$_2$ -- t$_3$ diagram for our sample of 31 H\,{\sc ii} regions. 
The filled circles are individual H\,{\sc ii} regions. The solid line is the 
linear best fit to those data derived through the least squares method. 
The dashed line is the t$_2$ -- t$_3$ relation derived in our previous 
study \citep{pvt06}.
}
\label{figure:tt-singl}
\end{figure}

Fig.~\ref{figure:tt-singl} shows the t$_2$ versus t$_3$ diagram for our sample 
of H\,{\sc ii} regions. 
Since the auroral line [OIII]$\lambda 4363$ is not detected in H\,{\sc ii} 
regions from our sample, the ff relation 
\begin{eqnarray}
\log {\rm R} & = & - 4.151- 3.118\,\log P + 
2.958 \, \log {\rm R}_{3}
\nonumber  \\
       & - & 0.680 \, (\log P)^2 , 
\label{equation:ffo}
\end{eqnarray}
derived in \citet{pilyuginetal06} was used to estimate the value of R. 
The filled circles are individual H\,{\sc ii} regions.
The solid line is the linear best fit to those data 
\begin{eqnarray}
t_2 =  0.746 \, (\pm 0.053) \times t_3  + 0.252 \, (\pm 0.039)
\label{equation:ttsingl}
\end{eqnarray}
derived through the least squares method. The mean value of residuals to 
this t$_2$ -- t$_3$ relation is 0.046. 

Recently we have derived a new $t_2$ -- $t_3$ relation based on the idea that 
the equation of the T$_{\rm e}$ method for O$^{++}$/H$^+$ applied to the O$^{++}$ 
zone and the equation for O$^{+}$/H$^+$ applied to the O$^{+}$ zone must result 
in the same value of the oxygen abundance \citep{pvt06}. The following 
$t_2$ -- $t_3$ relation has been derived: 
\begin{equation}
t_2 = 0.72 \, t_3 + 0.26 .
\label{equation:tt06}
\end{equation}
This relation is shown in Fig.~\ref{figure:tt-singl} by 
the dashed line. Inspection of Fig.~\ref{figure:tt-singl} as well the comparison 
between Eq.(\ref{equation:ttsingl}) and Eq.(\ref{equation:tt06}) shows that the 
two $t_2$ -- $t_3$ relations derived in different ways are close to each other. 

%----------------------------------------------------------------
\subsection{Two-dimensional (parametric) t$_2$ -- t$_3$ relation}
%----------------------------------------------------------------

\begin{figure}
\resizebox{1.00\hsize}{!}{\includegraphics[angle=000]{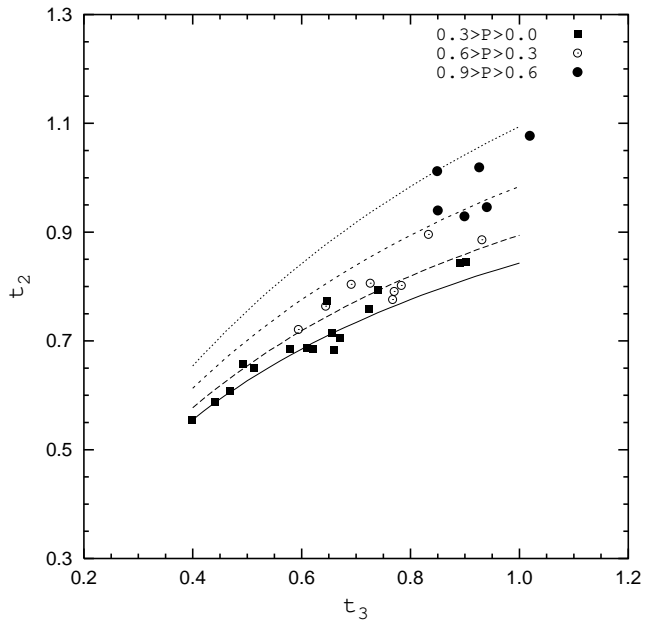}}
\caption{The t$_2$ -- t$_3$ diagram for our sample of H\,{\sc ii} regions. 
The filled squares are H\,{\sc ii} regions with 0 $<$ P $<$ 0.3. The open 
circles are those with 0.3 $<$ P $<$ 0.6. The filled circles are those with 
0.6 $<$ P $<$ 0.9. The relations corresponding to Eq.(\ref{equation:ttnit}) 
for different values of the excitation parameter are shown by the solid 
(P = 0.1), long-dashed  (P = 0.3), short-dashed  (P = 0.6), and dotted 
(P = 0.9) lines. }
\label{figure:t2-t3}
\end{figure}

Generally speaking, one might expect that there is no 
one-to-one correspondance between t$_2$ and t$_3$ temperatures, and that instead the 
t$_2$ -- t$_3$ relation is a function of the excitation parameter P. 
We now examine this possibility. Fig.~\ref{figure:t2-t3} shows the t$_2$ 
versus t$_3$ diagram for our sample of H\,{\sc ii} regions. The filled squares 
are H\,{\sc ii} regions with 0 $<$ P $<$ 0.3. The open circles are those with 
0.3 $<$ P $<$ 0.6. The filled circles are those with 0.6 $<$ P $<$ 0.9.
Close examination of Fig.~\ref{figure:t2-t3} suggests that the t$_2$ -- t$_3$ relationship 
is not unique but instead is dependent on and additional 
parameter -- the excitation parameter P. A fit to the data (31 data points) 
results in
\begin{eqnarray}
\frac{1}{t_2} & = &   0.410 \, (\pm 0.028) \times \frac{1}{t_3} 
- 0.344 \, (\pm 0.050) \times P 
\nonumber  \\
       & + & 0.818 \, (\pm 0.054) .
\label{equation:ttnit}
\end{eqnarray}
The relations corresponding to Eq.(\ref{equation:ttnit}) for various values of 
the excitation parameter are shown in Fig.~\ref{figure:t2-t3} 
by the solid (P = 0.1), long-dashed 
(P = 0.3), short-dashed (P = 0.6), and dotted (P = 0.9) lines. The mean value 
of residuals of this t$_2$ -- t$_3$ relation is 0.030, which is lower by a 
factor of $\sim$ 1.5 than the analogous value associated with the one-dimensional t$_2$ - t$_3$ relation. 

%====================================Fig  4    t2 - t3 relation

Thus, we have found evidence of that there is no one-to-one correspondence 
between t$_2$ and t$_3$ temperatures; instead the t$_2$ -- t$_3$ relation is 
dependent on the excitation parameter. The validity of the two-dimensional 
t$_2$ -- t$_3$ relation can be confirmed (or questioned) by the consideration 
of an additional sample of H\,{\sc ii} regions. 

%====================================Fig  5    t2 - t3 Sloan relation
\begin{figure}
\resizebox{1.00\hsize}{!}{\includegraphics[angle=000]{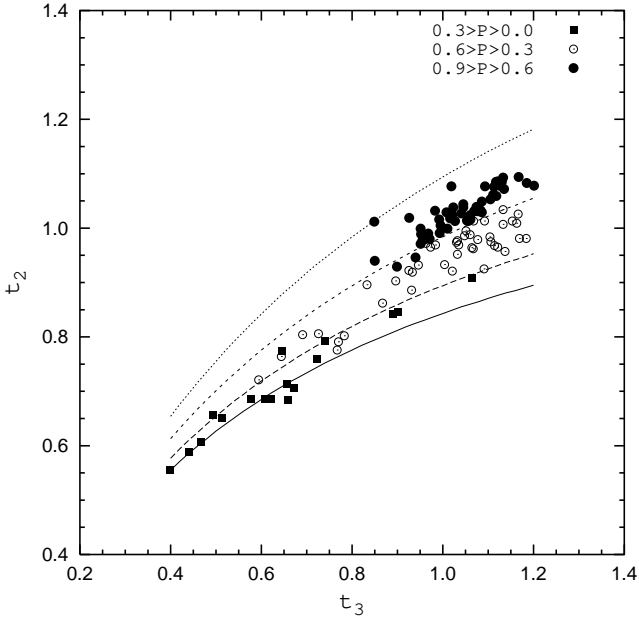}}
\caption{The t$_2$ -- t$_3$ diagram for extended sample of H\,{\sc ii} regions. 
The filled squares are H\,{\sc ii} regions with 0 $<$ P $<$ 0.3. The open 
circles are those with 0.3 $<$ P $<$ 0.6. The filled circles are those with 
0.6 $<$ P $<$ 0.9. The relations corresponding to Eq.(\ref{equation:tt}) for 
different values of the excitation parameter are shown by the solid (P = 0.1),  
long-dashed  (P = 0.3), short-dashed  (P = 0.6), and dotted (P = 0.9) lines. }
\label{figure:tt-slon}
\end{figure}

\citet{izotovetal04,izotovetal06} have extracted from the Data Release 3 of 
the Sloan Digital Sky Survey (SDSS) around 4500 spectra of H\,{\sc ii} 
regions with an [O\,{\sc iii}]~$\lambda 4363$  emission line detected at 
a level better than 1$\sigma$, and have carefully measured the line intensities in 
each spectrum. Yuri Izotov and Natalia Guseva have kindly provided me with the 
total list of their measurements, as only part of these have been published 
\citep{izotovetal04,izotovetal06}. It constitutes one of the largest and most
homogeneous data sets now available, being obtained and reduced in the same way.
The ff relation provides a way to select out the H\,{\sc ii} 
regions with high quality measurements \citep{pilyuginthuan05,pilyuginetal06};   
the discrepancy index D$_{\rm ff}$ allows us to eliminate low-quality 
measurements with large D$_{\rm ff}$, while retaining high-quality ones with 
small D$_{\rm ff}$. We have extracted a subsample of high-quality (with 
absolute value of D$_{\rm ff}$ less than 0.01) measurements of 
high-metallicity (12+log(O/H) $>$ 8.25) H\,{\sc ii} regions.  
The subsample of 80 H\,{\sc ii} regions meets that (very hard) precision 
criterion. The electron temperature t$_{3}$ in those H\,{\sc ii} regions is 
derived from the measured Q$_{\rm OIII}$ ratio, The electron temperature 
t$_{2}$ is derived from the Q$_{\rm NII}$ ratio given by Eq.(\ref{equation:ff}). 
Those data coupled with the data considered above form the total (extended) 
sample of 111 data points. 

Fig.~\ref{figure:tt-slon} shows the t$_2$ -- t$_3$ diagram for the total 
sample of H\,{\sc ii} regions. The filled squares are H\,{\sc ii} regions with 
0 $<$ P $<$ 0.3. The open circles are those with 0.3 $<$ P $<$ 0.6. The filled 
circles are those with 0.6 $<$ P $<$ 0.9. Inspection of Fig.~\ref{figure:tt-slon} 
confirms that the t$_2$ -- t$_3$ relationship is dependent on the excitation 
parameter P. A fit to those data (111 data points) gives 
\begin{eqnarray}
\frac{1}{t_2} & = &   0.418 \, (\pm 0.010) \times \frac{1}{t_3} 
- 0.348 \, (\pm 0.017) \times P 
\nonumber  \\
       & + & 0.806 \, (\pm 0.018)     .
\label{equation:tt}
\end{eqnarray}
The relations corresponding to Eq.(\ref{equation:tt}) for different values 
of the excitation parameter are shown in Fig.~\ref{figure:tt-slon} by the solid 
(P = 0.1), long-dashed (P = 0.3), short-dashed (P = 0.6), and dotted (P = 0.9) 
lines. Examination of Eq.(\ref{equation:ttnit}) and 
Eq.(\ref{equation:tt}) shows that the t$_2$ -- t$_3$ relations derived from the 
two different samples of H\,{\sc ii} regions are very similar and agree within 
the formal uncertainties. Thus the derived t$_2$ -- t$_3$ relation is rather 
robust. In the following, we will adopt as the t$_2$ -- t$_3$ relation 
\begin{eqnarray}
\frac{1}{t_2} & = &   0.41 \, \times \frac{1}{t_3} 
- 0.34 \, \times P  +  0.81     .
\label{equation:ttfin}
\end{eqnarray}

Thus, the consideration of an additional sample of H\,{\sc ii} regions strengthens 
the conclusion that the t$_2$ -- t$_3$ relation is two-dimensional or parametric. 

%--------------------------------------------
\subsection{Comparison with previous studies}
%--------------------------------------------

%====================================Fig  5    t2 - t3 Comparison
\begin{figure}
\resizebox{1.00\hsize}{!}{\includegraphics[angle=000]{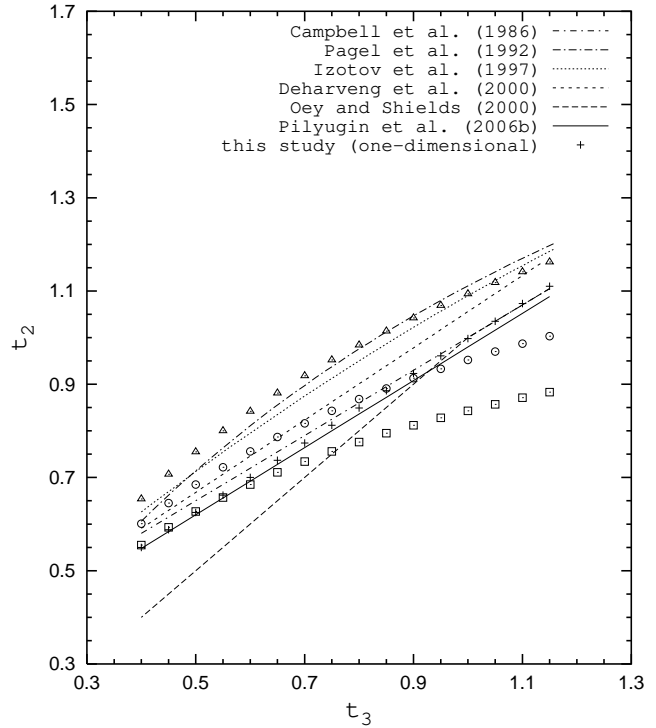}}
\caption{Comparison of the t$_2$ -- t$_3$ relation derived here with those 
derived by other investigators. The lines are t$_2$ -- t$_3$ relations derived 
in previous works. The one-dimensional t$_2$ -- t$_3$ relation derived here is 
shown by the plus signs. The two-dimensional (parametric) t$_2$ -- t$_3$ 
relation for different values of the excitation parameter are shown by the 
triangles (P = 0.9), the open circles (P = 0.5), and the open squares (P = 0.1). }
\label{figure:compar}
\end{figure}

Several versions of the $t_2$ -- $t_3$ relation have been proposed. 
A widely used relation is the one by \citet{campbelletal86} 
(see also \citet{garnett92}) based on the H\,{\sc ii} region models of 
\citet{stasinska82}. \citet{campbelletal86} has found that the $t_2$ -- $t_3$  
relationship can be parametrized as
\begin{equation}
t_2 = 0.7 \, t_3 + 0.3.
\label{equation:ttcam}
\end{equation}
Another relation has been proposed by \citet{pageletal92}, also based on 
H\,{\sc ii} region model calculations by \citet{stasinska90}, is: 
\begin{equation}
\frac{1}{t_2} = 0.5(\frac{1}{t_3} + 0.8).
\label{equation:ttpag}
\end{equation}
\citet{izotovetal97}, who also fitted the H\,{\sc ii} region models 
of \citet{stasinska90}, have proposed the following expression:
\begin{equation}
t_2 =  0.243 + 1.031\,t_3  - 0.184\,t_3^2  .
\label{equation:ttiz}
\end{equation}
Based on H\,{\sc ii} region model calculations by \citet{stasinskaschaerer97}),  
\citet{deharvengetal00} have suggested the following relation: 
\begin{equation}
t_2 = 0.775 \, t_3 + 0.281.
\label{equation:ttdeh}
\end{equation}
\citet{oeyshields00} have found that the Campbell et al. relation 
is reasonable for $t_{3}$ $>$ 1.0. However at lower temperatures, 
the models are more consistent with an isothermal nebula. They consequently
adopted the formulation:
\begin{eqnarray}
t_2   = & 0.7 \, t_3 + 0.3, &  \;\; t_3 > 1.0   \nonumber  \\
        & t_3,              &  \;\; t_3 < 1.0 .
\label{equation:ttoey}
\end{eqnarray}

We now compare the t$_2$ -- t$_3$ relation obtained here with those obtained 
by other authors. Since our t$_2$ -- t$_3$ relation is derived for cool 
high-metallicity H\,{\sc ii} regions, the high-temperature low-metallicity 
part of relation is not considered here. The lines in Fig.~\ref{figure:compar} 
are t$_2$ -- t$_3$ relations from 
\citet{campbelletal86,pageletal92,izotovetal97,deharvengetal00,oeyshields00,pvt06}. 
Our one-dimensional t$_2$ -- t$_3$ relation is shown by the plus signs.
Our two-dimensional t$_2$ -- t$_3$ relations for different values 
of the excitation parameter are shown by the triangles (P = 0.9),  
the open circles (P = 0.5), and the open squares (P = 0.1). 
Examination of Fig.~\ref{figure:compar} shows that the t$_2$ -- t$_3$ relations 
of  \citet{pageletal92} and \citet{izotovetal97} are relatively close to our 
two-dimensional relation for high excitation H\,{\sc ii} regions, while 
at low temperatures, the t$_2$ -- t$_3$ relations of  
\citet{campbelletal86} and \citet{deharvengetal00} are relatively close to our 
two-dimensional relation for moderate excitation H\,{\sc ii} regions. 

\section{Conclusions}

A relationship between the ratio of nebular to auroral nitrogen line 
intensities, which is an indicator of the electron temperature t$_2$ in 
the O$^{+}$ zone of H\,{\sc ii} regions, and nebular oxygen line intensities
in spectra of high-metallicity H\,{\sc ii} regions was derived. 
Since the oxygen nebular lines are strong and, consequently, are easily 
observable, the derived relation coupled, with the ff relation derived 
in our previous studies \citep{ff,pilyuginetal06}, solve the problem 
of the determination of the electron temperatures t$_2$ and t$_3$ 
in high-metallicity H\,{\sc ii} regions 
where faint auroral lines are not detected.
It should be emphasized that the relation obtained here and the ff relation 
are purely empirical relations in the sence that they are relations between 
directly measured values. Consequently, there is no assumptions at the 
base of those relations.

The derived relation and the ff relation confirm our idea 
that the diagnostic line ratios, which are indicators of the electron 
temperatures, can be expressed in terms of the strong oxygen nebular lines. 
This, in turn,  confirms the basic assumption of the 
``empirical'' method, proposed by \citet{pageletal79} a quarter of a century 
ago, that the oxygen abundance in H\,{\sc ii} regions can be estimated  from 
strong oxygen line measurements only. 

The relation between electron temperatures in the O$^{++}$ and O$^{+}$ zones 
in high-metallicity H\,{\sc ii} regions was investigated. 
It was found that there is no one-to-one correspondance between t$_2$ 
and t$_3$ temperatures. Instead the t$_2$ -- t$_3$ relation is dependent on 
the excitation parameter. 

\subsection*{Acknowledgments}

   I thank Yuri Izotov and Natalia Guseva for providing me with their 
measurements of the line intensities in spectra of H\,{\sc ii} regions extracted 
from the Data Release 3 of the Sloan Digital Sky Survey (SDSS), the majority 
of which are unpublished. 
   I am grateful to R.B.C. Henry for constructive comments as well as improving 
the English text.
   I thank G.Stasi\'{n}ska for usefull discussion. 
   I am gratefull to an anonymous referee for useful comments that improved the 
   presentation of the paper.
   The author acknowledges the work of the Sloan Digital Sky
Survey (SDSS) team. Funding for the SDSS has been provided by the
Alfred P. Sloan Foundation, the Participating Institutions, the National
Aeronautics and Space Administration, the National Science Foundation, the
U.S. Department of Energy, the Japanese Monbukagakusho, and the Max Planck
Society. The SDSS Web site is http://www.sdss.org/.
The SDSS is managed by the Astrophysical Research Consortium (ARC) for
the Participating Institutions. The Participating Institutions are The
University of Chicago, Fermilab, the Institute for Advanced Study, the Japan
Participation Group, The Johns Hopkins University, the Korean Scientist Group,
Los Alamos National Laboratory, the Max-Planck-Institute for Astronomy (MPIA),
the Max-Planck-Institute for Astrophysics (MPA), New Mexico State University,
University of Pittsburgh, University of Portsmouth, Princeton University, the
United States Naval Observatory, and the University of Washington.

\end{document}